\documentclass[12pt]{iopart}
\pdfoutput=1

\usepackage{iopams}
\usepackage{hyperref}
\usepackage{amsthm}
\usepackage{amssymb}
\usepackage{graphicx}
\usepackage[dvipsnames]{xcolor}
\usepackage{tabularx}
\usepackage{xfrac}
\usepackage{comment}

\begin{document}
\title[Hamiltonian Engineering with Constrained Optimization]{Hamiltonian Engineering with Constrained Optimization for Quantum Sensing and Control}
\author{Michael F. O'Keeffe$^{1,4}$, Lior Horesh$^{2,4}$, John F. Barry$^{1}$, Danielle A. Braje$^{1,4}$, Isaac L. Chuang$^{3,4}$}
\address{$^1$ Quantum Information and Integrated Nanosystems Group, MIT Lincoln Laboratory, Lexington, MA 02421, USA}
\address{$^2$ AI Science, IBM T.J. Watson Research Center, Yorktown Heights, NY 10598, USA}
\address{$^3$ Department of Physics, Department of Electrical Engineering and Computer Science, and Research Laboratory of Electronics, Massachusetts Institute of Technology, Cambridge, MA 02139, USA}
\address{$^4$ MIT-IBM Watson AI Lab, Cambridge, MA 02142, USA}

\ead{michael.okeeffe@ll.mit.edu}

\begin{abstract}
While quantum devices rely on interactions between constituent subsystems and with their environment to operate, native interactions alone often fail to deliver targeted performance. 
Coherent pulsed control provides the ability to tailor effective interactions, known as Hamiltonian engineering.  
We propose a Hamiltonian engineering method that maximizes desired interactions while mitigating deleterious ones by conducting a pulse sequence search using constrained optimization.  
The optimization formulation incorporates pulse sequence length and cardinality penalties consistent with linear or integer programming.  
We apply the general technique to magnetometry with solid state spin ensembles in which inhomogeneous interactions between sensing spins limit coherence. 
Defining figures of merit for broadband Ramsey magnetometry, we present novel pulse sequences which outperform known techniques for homonuclear spin decoupling in both spin-1/2 and spin-1 systems.
When applied to nitrogen vacancy (NV) centers in diamond, this scheme partially preserves the Zeeman interaction while zeroing dipolar coupling between negatively charged NV$^{\text -}$ centers. 
Such a scheme is of interest for NV$^\text{-}$ magnetometers which have reached the NV$^\text{-}$-NV$^\text{-}$ coupling limit. 
We discuss experimental implementation in NV ensembles, as well as applicability of the current approach to more general spin bath decoupling and superconducting qubit control.  
\end{abstract}

\maketitle


\section{Introduction}

From quantum computing to sensing, both the strength and fragility of the underlying quantum systems stem from interactions.  
Coupling enables gate operations and metrology, but also introduces crosstalk and decoherence.  
As quantum technologies mature, the demands of tailoring properties and dynamics increase.  
Hamiltonian engineering describes a family of classical control techniques on quantum systems to achieve a desired state, process, or observable behavior~\cite{Schirmer2007}.  
This encompasses decoupling to eliminate or reduce unwanted interactions~\cite{Ajoy2019,Choi2017}, optimal control using numerical optimization for maximizing the fidelity of quantum computing operations given experimental limitations~\cite{Khaneja2005,Rebentrost2009,Schulte-Herbruggen2011,Nobauer2015,Khodjasteh2012}, and Hamiltonian simulation~\cite{Lloyd1996,Ajoy2013,Hayes2014} of models for exploring quantum phenomena not otherwise immediately accessible in experiment~\cite{Salathe2015,Schreiber2015}.  

Solid-state quantum systems are emerging as viable candidates for high-performance sensing~\cite{Degen2017}, both filling existing technology gaps, such as stable vector magnetometry~\cite{Clevenson2018,Schloss2018}, and opening new capabilities in areas like high-spatial-resolution sensing~\cite{Taylor2008,Balasubramanian2008,Aslam2017}.  
Ensemble magnetometry with negatively charged nitrogen vacancy (NV$^{\text -}$) centers in diamond~\cite{Jensen2017} has demonstrated sensitivities of 15 pT$/\sqrt{\text{Hz}}$ for DC~\cite{Barry2016} and 0.9 pT$/\sqrt{\text{Hz}}$ for AC~\cite{Wolf2015} sensing.  In the latter, the application of dynamical decoupling consistent with narrowband magnetometry~\cite{Pham2012,Wang2012} suppresses low frequency fluctuations and inhomogeneity, yielding improved performance.  Still, $T_2^*$ dephasing times and $T_2$ coherence times currently remain orders of magnitude shorter than $T_1$ spin relaxation times~\cite{Myers2017}, leaving much room for improvement to reach the spin relaxation limited coherence time $T_2 = 2 T_1$.  

The spin-projection-limited sensitivity of an ensemble magnetometer consisting of $N$ spins can be intuitively understood by~\cite{Budker2007}
\begin{equation}
\eta = \frac{1}{\gamma C\sqrt{N\tau}}
\label{eq:sensitivity}
\end{equation}
where $\gamma$ is the gyromagnetic ratio, $C$ the measurement contrast, and $\tau$ the interrogation time during which the spins precess, typically on the order of the relevant coherence time, i.e. $T_2^*$ for DC magnetometry and $T_2$ for AC magnetometry.  
To date, most spin environment engineering for solid-state ensembles has focused on extending NV$^{\text -}$ coherence time by synthesizing diamond to eliminate other spin impurities, which commonly include other electronic defects (substitutional nitrogen atoms known as P1 centers, neutrally charged NV$^0$, divacancies, NVH$^\text{-}$)~\cite{Acosta2009}, as well as nuclear spin species ($^{13}$C)~\cite{Childress2006,Stanwix2010}.  
Although isotope engineering is efficient to remove nuclear spin dephasing~\cite{Balasubramanian2009}, the interdependence between NV center formation and the incorporation of a number of spin impurity species makes eliminating all spin impurities via diamond growth intractable.  
Various broadening mechanisms from paramagnetic defects can be mitigated by driving the environment spins~\cite{deLange2012,Knowles2013} while inhomogeneous broadening mechanisms like strain variation and temperature fluctuation can be mitigated by double quantum~\cite{Fang2013,Mamin2014}.  
Used independently, these techniques help to identify sources of dephasing; in combination, they can result in over an order of magnitude improvement in $T_2^*$~\cite{Bauch2018}.  NV$^{\text -}$-NV$^{\text -}$ dipolar interactions will likely ultimately limit NV$^{\text -}$ coherence time~\cite{Bar-Gill2013}.  

Inhomogeneous interactions between spins in an ensemble limit coherence~\cite{Abragam1961}, with each spin experiencing a different local magnetic field $B_{\rm loc}\sim\mu/r^3$ due to a neighboring magnetic moment $\mu$ a distance $r$ away.  
A rough estimate for the dephasing time $T_2^* \simeq 1/\gamma B_{\rm loc}$ shows that it varies inversely to the density of spins $n = N/V$.  
While increasing the spin density for a fixed sensor volume $V \sim r^3$ appears to improve (decrease) the sensitivity in~\eref{eq:sensitivity}, its effect on coherence time, and thus sensing time $\tau$, must also be accounted for.  

\begin{figure*}
\includegraphics[width=\textwidth, trim={0 2.5cm 0 1cm},clip]{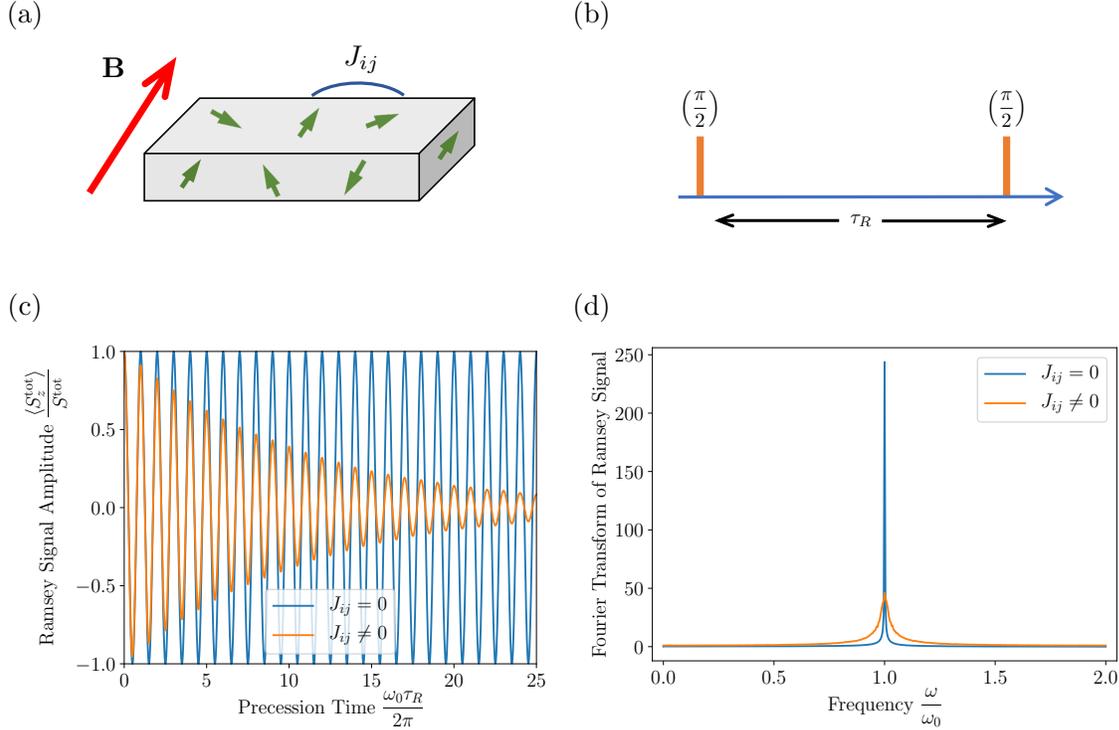}
\caption{\textbf{NV ensemble Ramsey magnetometry.} (a) Schematic of a spin ensemble subject to a global magnetic field $\mathbf B$ with pairwise interaction strengths $J_{ij}$ between spins.  (b) Ramsey pulse sequence.  (c) Free induction signal and (d) corresponding spectrum of a simulated Ramsey experiment with no interactions (blue), and an average over many interaction strengths (orange), with Gaussian linewidths drawn from a distribution of coupling strengths \eref{eq:linewidth_distribution} representative of an ensemble.  Without interactions, the free induction signal oscillates sinusoidally producing a sharp spectral line at the Larmor frequency $\omega_0 = \gamma B_z$, whereas inhomogeneous interactions cause free induction decay and a broadening of the spectral line.  
\label{fig:ensemble_schematic_FID_spectrum}}
\end{figure*}

Ramsey magnetometry entails preparing a superposition state, letting it undergo free evolution for some time $\tau_R$, then rotating the state to a convenient basis for readout, with the typical scheme shown in \fref{fig:ensemble_schematic_FID_spectrum} (b).  For an ``ideal'' ensemble with no interactions between spins or with the environment, the Ramsey signal is a sinusoid with angular frequency $\omega_0 = \gamma B$, whereas inhomogeneous interactions or coupling to the environment, typical of solid state systems, produces free induction decay, or equivalently a broadened spectral line, shown in \fref{fig:ensemble_schematic_FID_spectrum} (c) and (d), respectively.

Solid state spins are subject to a large number of anisotropic interactions, including chemical shifts, dipolar, and quadrupolar, which typically leads to broad spectral lines in nuclear magnetic resonance (NMR) and electron spin resonance.  A line narrowing workhorse of solid state NMR, the WHH-4 pulse sequence~\cite{Waugh1968} decouples homonuclear dipolar interactions to leading order in spin-1/2 systems, and when repeated in reverse order (MREV-8) also compensates for pulse imperfections~\cite{Mansfield1973, Rhim1973, Rhim1974}.  Choi, Yao, and Lukin recently discovered a six pulse sequence~\cite{Choi2017}, which we refer to as CYL-6, that achieves dipolar decoupling in spin-1 ensembles, effectively generalizing WHH-4.  Drawing motivation from this work, we evaluate both sequences in the context of Ramsey magnetometry, and find that neither sequence produces single frequency observable response.   Such an exercise makes clear the need to identify both desirable (the magnetic field to be detected) and undesirable (inhomogeneous line broadening) interactions and the corresponding terms in the Hamiltonian.    

Given a system's Hamiltonian with desirable and undesirable terms, the goal of Hamiltonian engineering is to find a set of unitary operators that transforms the original Hamiltonian to a target Hamiltonian containing only desirable terms with maximal strength.  
In this paper, we explicitly formulate Hamiltonian engineering as a quantum pulse sequence search problem with conditions for achievability and optimality (\sref{sec:ham_eng_as_qps_search}).  
We cast the pulse sequence search as constrained optimization in which the search criteria are formulated as constraints and the objective function incorporates sequence length and cardinality regularization suitable for linear or integer programming frameworks (\sref{sec:ps_search_as_constr_opt}).  
As a concrete example, we choose the goal of decoupling dipolar interactions in a spin ensemble used for broadband magnetometry (\sref{sec:dip_decoup_for_broad_mag}).  
The undesirable term is the inhomogeneous dipolar interaction, and the desirable term is a Zeeman term from a global magnetic field of interest.  
Success criteria include a clean, single frequency Ramsey magnetometry signal, and the strength, or frequency scaling, of the resulting oscillations.  
We present and analyze pulse sequences for qubit and qutrit systems that effectively average the dipolar interaction term in the Hamiltonian to zero, even with inhomogeneous interaction strengths, while preserving the desired Zeeman term with strength multiplied by 1/3.  
Because these sequences can be applied during the free evolution interval of a Ramsey experiment and preserve the sinusoidal dependence of the observed signal on the free evolution time, albeit with a frequency scaled by some constant, we name this family of sequences ``Homonuclear Ramsey decoupling'' (HoRD), and adopt the naming convention HoRD-[qudit type]-[number of pulses].
\Tref{tab:seq} compares this new family of pulse sequences with existing dipolar decoupling sequences for qubits and qutrits. 
We discuss considerations for experimental demonstration of the HoRD sequence for spin-1 systems in NV ensembles and, with general techniques germane to a variety of systems, point towards a number of possibilities for future work (\sref{sec:discussion}) prior to concluding in~\sref{sec:conclusion}.  

\begin{table}
\caption{Dipolar decoupling pulse sequence characteristics\label{tab:seq}}
\footnotesize
\begin{tabular}{p{0.16\textwidth}p{0.08\textwidth}p{0.10\textwidth}p{0.12\textwidth}p{0.4\textwidth}}
\br
Name & 
  Pulses & 
  Clean Zeeman & 
  Zeeman Strength &
  Notes\\ 
\mr
\textit{Spin-$\frac{1}{2}$ qubits} \\
WHH-4~\cite{Waugh1968} & 
  4 & 
    No & 
  $1/\sqrt{3}$ & 
  Original qubit dipolar decopuling sequence. Found analytically using average Hamiltonian theory. Simple pulses. Symmetric. \\
HoRD-qubit-5& 
  5 &
    Yes & 
  $1/3$ &
  Qubit dipolar decoupling sequence that leaves Zeeman term intact, producing a single-frequency Ramsey signal. Found analytically using average Hamiltonian theory.  Composite pulses. Symmetric. \\
\textit{Spin-1 qutrits} \\
CYL-6~\cite{Choi2017} &
  6 & 
    No & 
  $1/\sqrt{6}$ & 
  Qutrit dipolar decoupling sequence, found with linear programming with search set $\{x_1 x_2 x_3 x_4 | x \in \mathcal E\}$ where $\mathcal E$ is set of all $x$ and $y$ $\pi$ and $\pi/2$ pulses between pairs of levels.  \\
HoZD-qutrit-12 & 
  12 & 
    No & 
  0 & 
  Qutrit dipolar decoupling sequence that also cancels Zeeman term. Contains 6 pairs of terms that map Zeeman term to $\pm \lambda_i$, $i=1,\ldots,6$. Search set composed of $\{x_1 x_2 x_3 | x \in \mathcal C_i\}$ where $\mathcal C_i$ is set of all qubit Clifford operators for subsystem $i$.  \\
HoRD-qutrit-8 &
    8 & 
    Yes & 
  1/3 & 
  Qutrit dipolar decoupling.  Found by replacing four terms of HoZD-qutrit-12 with their conjugates, then searching for global unitary transformation that produces desired Zeeman term. Consistent with single quantum or double quantum magnetometry.\\
\br
\end{tabular}
\end{table}

\section{Hamiltonian engineering as a quantum pulse sequence search problem
\label{sec:ham_eng_as_qps_search}}

Hamiltonian engineering entails finding a set of operations, realized with a sequence of pulses, 
that transform a given Hamiltonian into a target Hamiltonian.  In this section, we review average Hamiltonian theory for periodically driven systems.  We then make the quantum pulse 
sequence search problem definition concrete, first abstractly, then explicitly, by using the traditional linear 
algebraic framework employed in quantum mechanics, and then by defining a convenient ``Pauli projection'' 
representation of the problem.  

\subsection{Average Hamiltonian theory}

The state of quantum systems can be manipulated with pulsed control, through which some set of  quantum operations can be performed.  
Average Hamiltonian theory~\cite{Haeberlen1968} provides a mathematical formalism to evaluate the time evolution of a system subject to a pulse sequence.  
Originally developed for NMR, these techniques have found renewed interest in the context of quantum control and computation~\cite{Vandersypen2005}.  

We capture the effect of intermittent driving with short ``delta function" pulses $P_k$ on the system separated by time intervals $\tau_k$ of free evolution during the $k^{\rm th}$ interval.  The time evolution operator 
\begin{equation}
    U(t_c) = e^{-iH\tau_n}P_n \cdots e^{-iH\tau_1}P_1 e^{-iH\tau_0}P_0
    \label{eq:U_tc_pulses}
\end{equation}
propagates the state of the system over a cycle time $t_c = \sum_{k=0}^n \tau_k$.  
The pulses transform the Hamiltonian  during the $k^{\rm th}$ interval as $\tilde H_k = U_k^\dagger H U_k$, where $U_k = P_k \cdots P_0$ or equivalently $P_k = U_k U_{k-1}^\dagger$, giving an equivalent form of the time evolution operator,
\begin{equation}
    U(t_c) = e^{-i\tilde H_n\tau_n} \cdots e^{-i\tilde H_1\tau_1} e^{-i\tilde H_0\tau_0}.
    \label{eq:U_tc_Hk}
\end{equation}

For a periodic time-dependent Hamiltonian $H(t) = H(t+t_c)$ where observation is stroboscopic and coordinated with the period $t_c$, average Hamiltonian theory~\cite{Haeberlen1968} shows that the time evolution operator can be expressed in terms of a time-independent average Hamiltonian $\bar H$
\begin{equation}
    U(t_c) = \exp(-i\bar{H}t_c).
    \label{eq:U_tc_avg}
\end{equation}  
Using the Magnus expansion~\cite{Magnus1954}, the average Hamiltonian
\begin{equation}
    \bar H = \bar H^{(0)} + \bar H^{(1)} + \bar H^{(2)} + \ldots
    \label{eq:avg_ham}
\end{equation}
has leading order contributions 
\begin{equation}
    \label{eq:avg_hamiltonian_0}
    \bar H^{(0)} = \frac{1}{t_c} \sum_{k=0}^n \tilde H_k \tau_k,\quad \tilde H_k = U_k^\dagger H U_k
\end{equation}
with the next order 
\begin{equation}
    \label{eq:avg_hamiltonian_1}
    \bar H^{(1)} = \frac{-i}{2 t_c} \sum_{k=0}^n\sum_{l<k} [\tilde H_k \tau_k, \tilde H_l \tau_l] 
    \,
\end{equation}
and higher orders involving commutators.  
$\bar H^{(0)}$ provides the quantity of interest for Hamiltonian engineering.  Higher order terms in the Magnus expansion vary as $(\|H\|t_c)^n/n!$, where the operator norm of $H$ gives the largest eigenvalue, so making the pulses and intervals short compared to the fastest timescale of the Hamiltonian generally yields better performance.  
Zeroing higher order terms, which is generally desirable, can also be achieved by symmetrizing a pulse sequence that gives the desired leading order behavior~\cite{Mansfield1973, Rhim1973, Rhim1974}, or incorporating higher order terms in the Magnus expansion into the optimization goal.  
For periodically driven systems, the so-called Floquet time evolution operator satisfies \begin{equation}
U(nt_c) = [U(t_c)]^n,
\label{eq:U_floquet}
\end{equation}
a property we will use for efficient time dependent simulation of the pulse sequences discussed in \sref{sec:dip_decoup_for_broad_mag}.  

\subsection{The search criteria and the search problem\label{subsec:search_crit}}

We first establish the mathematical setting for the Hamiltonian $H$, and unitary operators $U$,
then define the search criteria.  $H$ and $U$ are linear operators acting within a Hilbert space, a linear vector space
with well-defined inner product.  The operators are 
commonly defined in terms of the tensor product of the constituent
quantum elements of the system.  Each constituent element is typically
a $d$-dimensional quantum system (e.g. qubit, with $d=2$, or qutrit,
with $d=3$), with Hilbert space $SU(d)$.  A system of $n$ elements
will thus have $H$ and $U$ be $d^n \times d^n$ matrices.  For example,
operators in a system of two coupled qubits are represented by
$4\times 4$ matrices, and for two coupled qutrits, $9 \times 9$.  
$H$ and $U$ are constrained to be Hermitian and unitary, respectively.

$U$ is drawn from a set of $\{ U_j \}$ which are physically
realizable, and $H$ is typically fixed, but composed of three parts:
\begin{equation}	H = H_{\rm sys} + H_{\rm env} + H_{\rm sig}
\,.
\end{equation}
$H_{\rm sys}$ describes the system's own internal dynamics; $H_{\rm  env}$ 
describes couplings to the environment, and $H_{\rm sig}$
describes couplings to an external signal, including classical control and quantities which it may be desired to
measure.

The goal is typically to zero out certain terms (call these $H_0$), while keeping
other terms intact (call these $H_1$).  For this purpose, let us regroup $H$ as:
\begin{eqnarray}
	H = H_0 + H_1
\,.
\end{eqnarray}
This identification of $H_0$ and $H_1$ is the specification of
${\mathcal G}$, the goal criterion.  The zeroing out is accomplished by applying sequences of unitary rotations $U_k$ to the system, such that the effective first order ``average Hamiltonian'' (see \eref{eq:avg_hamiltonian_0} and surrounding discussion) is:
\begin{eqnarray}
	H_{\rm avg} = \frac{1}{\sum_k w_k} \sum_k w_k U_k^\dagger H U_k
\,,
\end{eqnarray}
where $w_k$, the weight for unitary $U_k$, is proportional to the amount
of time spent during evolution in the reference frame determined by
$U_k$, known as the toggling frame.  The weights $w_k$ are positive and nonzero, and typically are
integers, but in principle they could be real numbers.

The quantum pulse sequence search problem is thus summarized as follows.
Given $H = H_0 + H_1$ and a set of unitaries ${\mathcal U} = \{U_j\}$,
\begin{itemize}
\item {\bf Achievability}:  Does there exist any subset ${\mathcal U}_{\rm seq} \subset {\mathcal U}$ of 
  available unitaries, and weights $w_k$ such that
  \begin{eqnarray}
	\frac{1}{\sum_k w_k} \sum_{U_k\in {\mathcal U}_{\rm seq}}  w_k U_k^\dagger \left[ H_0 + H_1 \right] U_k = \beta H_1
\label{eq:searchgoal}
  \,,
  \end{eqnarray}
  where $\beta$ is bounded away from zero by a constant?
\item {\bf Optimality}: What is the shortest sequence such that the
  above equation holds, with the largest $\beta$ (ideally $\beta=1$),
  smallest size $|{\mathcal U}_{\rm seq}|$, and smallest total weight
  $\sum_k w_k$?
\end{itemize}
Note that this formalism is completely general to Hamiltonian engineering, and not limited to interaction decoupling.  For example, to obtain and maximize a desired interaction term $H_2$ that does not occur in the original Hamiltonian $H$, simply replace $H_1$ with $H_2$ on the right hand side of \eref{eq:searchgoal}.  Ref.~\cite{Choi2017} presents necessary and sufficient conditions for both decoupling and engineered interactions.  

\subsection{The Pauli projection representation
\label{subsec:pauli_projection}}

The operators $H$ and $U$ arise from elementary dynamics of constituent 
subsystems. Thus, it is highly convenient to represent all these operators in
terms of the $d^2-1$ generators $\{\sigma_k \}$ of $SU(d)$.  Let these
be indexed from $k=1$ to $k=d$, and define $\sigma_0$ as the
identity. Each $\sigma_k$ is a $d\times d$ matrix; for qubits, they are the
Pauli matrices, and for qutrits they are commonly taken to be the
Gell-Mann matrices.  Explicit representations as well as useful properties are given in \nameref{sec:appendix_A}.  These are all Hermitian, i.e. $\sigma_k^\dagger =
(\sigma_k^T)^* = \sigma_k$. Except for $\sigma_0$, these are
traceless matrices, and they satisfy a trace orthogonality condition,
\begin{eqnarray}
    {\rm Tr}\left( \sigma_j \sigma_k \right) = c_d \delta_{ij}
\,.
\end{eqnarray}
We shall refer to the $\sigma_k$ matrices as the generalized Pauli matrices. 
In addition to the trace orthogonality condition, an exceptionally
useful fact is that they form a linear basis for any $d\times d$
matrix, i.e. any such matrix $M$ can be expressed as a linear combination 
of the $\sigma_k$:
\begin{eqnarray}
	M = \sum_k c_k \sigma_k
\,,
\end{eqnarray}
and because of the trace orthogonality condition,
\begin{eqnarray}
	c_k = \frac{1}{\alpha_d} {\rm Tr}( M\sigma_k )
\,,
\end{eqnarray}
where the normalization factor $\alpha_d$ depends on how the $\sigma_k$ are
defined.  For Hermitian matrices such as $H$, $c_k \in \mathbb{R}$, whereas 
for general $M$, including e.g. $U$, $c_k \in \mathbb{C}$

Let the full quantum system comprise $n$ constituent subsystems.  Then
the $n$-fold tensor products of the generalized Pauli matrices
\begin{eqnarray}
          S_{i_1 i_2\cdots i_n} = \sigma_{i_1} \otimes \sigma_{i_2} \otimes \cdots \otimes \sigma_{i_{n}}
\end{eqnarray}
form a linear basis for the operator space, giving a nice way to express $H$, as
\begin{eqnarray}
	H = \sum_{i_1 i_2\cdots i_n} c_{i_1 i_2\cdots i_n} S_{i_1 i_2\cdots i_n}
\,.
\end{eqnarray}
We define the Pauli projection of $H$,
\begin{eqnarray}
    {\mathcal P}(H) &=& \left\{ \frac{1}{\alpha_d^n} {\rm Tr}( H S_{i_1 i_2\cdots i_n} ) \ \Bigg|\ \forall i_1 i_2\cdots i_n \right\}
\\
     &=& \left\{ c_{i_1 i_2\cdots i_n} \right\}
\\
     &\equiv& \boldsymbol{c}_H
\,.
\end{eqnarray}    
This isomorphic map represents $H$ using a set of $d^{2n}$ real numbers, 
which we shall find convenient to vectorize as $\boldsymbol{c}_H$, and call the 
Pauli projection coefficients.   

\subsection{Matrix-vector formulation of the search problem}

The first-order search problem, defined in \eref{eq:searchgoal}, is
conveniently re-expressed using the Pauli projection, by applying the map to
each term in the sum, to define Pauli projection coefficients for each of the 
$U_k$ transformed Hamiltonian terms:
\begin{eqnarray}
	\boldsymbol{c}_k^{\,0} &=& {\mathcal P}(U_k^\dagger H_0 U_k)
\\
	\boldsymbol{c}_k^{\,1} &=& {\mathcal P}(U_k^\dagger H_1 U_k)
\,.
\end{eqnarray}
Recall that the goal is to average away $H_0$, but leave $H_1$ intact.  Thus, 
in terms of these vectors, the search problem becomes two sets of equations:
\begin{eqnarray}
	  \sum_k  w_k \boldsymbol{c}_k^{\,0} &=& 0
\label{eq:search_zero}
\\
	  \sum_k  w_k \boldsymbol{c}_k^{\,1} &=& \beta' \boldsymbol{c}_{H_1}
\label{eq:search_nonzero}
\,,
\end{eqnarray}
where $\beta' = \beta {\sum_k w_k}$.
The matrix-vector formulation relies on our choice of solving the first order achievability of an average Hamiltonian where linear dependence on the pulse sequence dominates.  This is well justified if the duration of each pulse and intervals between pulses are sufficiently short.  

\section{Pulse sequence search as constrained optimization
\label{sec:ps_search_as_constr_opt}}

Brute force search checks every possible solution, and is guaranteed to find a solution if it exists in the search space.  The computational complexity is linear in the number of possible solutions.  Because sequences made of composite pulses contain combinations of available pulses, the number of possible solutions scales combinatorially with the number of available pulses.
We prune operators that transform the system Hamiltonian in the same way, and provide two relevant examples to establish a sense of scale.  The 24 qubit Clifford operators, defined in Section~\ref{subsec:qubit_ensembles}, only produce 6 unique mappings of the Hamiltonian describing dipolar interacting pins (spin-1/2) subject to a global magnetic field, \eref{eq:H}, allowing us to prune the search set from 24 to 6 operators.  
The 13,824 Clifford operators between sublevels in a qutrit system produce 558 unique mappings of the corresponding Hamiltonian for spin-1.  While this reduces the number of elements in the search table by an order of magnitude, there are over 10$^{13}$ and 10$^{24}$ combinations for 6 and 12 pulse sequences, respectively.
To overcome the potentially prohibitive cost of a brute-force search, we cast the pulse sequence search problem as constrained optimization.  Concretely, we formulate the search problem, \eref{eq:search_zero}-\eref{eq:search_nonzero}, as linear constraints, and impose sequence length and cardinality penalties in the linear objective function.  

\subsection{Constraint formulation}

The pulse sequence search problem, as defined by
\eref{eq:search_zero}-\eref{eq:search_nonzero} can immediately be
cast in terms of the standard framework for constrained optimization
using linear programming.  In that framework, we wish to find a vector $\boldsymbol{x}$ such that:
\begin{eqnarray}
	\boldsymbol{A}_{\rm eq} \boldsymbol{x} &=& \boldsymbol{b}_{\rm eq}
\label{eq:linprog_eq}
\\
	\boldsymbol{A}_{\rm ub} \boldsymbol{x} &\leq& \boldsymbol{b}_{\rm ub}
\label{eq:linprog_ub}
\,,
\end{eqnarray}
where $\boldsymbol{A}_{\rm eq}$ and $\boldsymbol{A}_{\rm ub}$ are matrices, and together with
vectors $\boldsymbol{b}_{\rm eq}$ and $\boldsymbol{b}_{\rm ub}$, specify the
optimization problem.

\def\nvec#1{{\mathcal N}\left(\boldsymbol{c}_{#1}^{\,1}\right)}
\def\zvec#1{{\mathcal Z}\left(\boldsymbol{c}_{#1}^{\,0}\right)}

For our search problem, from \eref{eq:search_nonzero}, suppose ${\mathcal N}$
projects into the subspace of $\boldsymbol{c}_{H_1}$ which has nonzero
coefficients, such that
${\mathcal N}(\boldsymbol{c}_{H_1}) > 0$.  Let the length of this projected vector be $m_{\rm ub}$.  Then we may choose
$\boldsymbol{b}_{\rm ub} = -{\mathcal N}(\boldsymbol{c}_{H_1})$, and $\boldsymbol{A}_{\rm ub}$ is the $m_{\rm ub} \times N_U$ matrix
formed by taking $\nvec{k}$ as the $k^{\rm th}$ column, i.e.
\begin{eqnarray}
    \boldsymbol{A}_{\rm ub} = - \left[ \begin{array}{c|c|c|c}
      & & & \\
      \nvec{0} & \nvec{1} & \cdots & \nvec{N_U} \\
      & & & \\
    \end{array} \right]
\,,
\end{eqnarray}
where $N_U = |{\mathcal U}|$.  Note the overall minus sign on $\boldsymbol{A}_{\rm ub}$, 
and the minus sign on $\boldsymbol{b}_{\rm ub}$: these are present
because we actually have a lower bound, but the standard form of the
optimization equations expects an upper bound.  Similarly, we may define
${\mathcal Z}$ as the projector into the subspace of Pauli projection
coefficients expected to be zero (e.g. ${\mathcal Z} = I - {\mathcal
  N}$ for suitable definition of $I$).  Let the length of ${\mathcal
  Z}(\boldsymbol{c}_{H_0})$ be $m_{\rm eq} = d^{2n} - m_{\rm ub}$.  Thus,
from \eref{eq:search_zero} we have that $\boldsymbol{b}_{\rm eq} = 0$,
and $\boldsymbol{A}_{\rm eq}$ is the $m_{\rm eq}\times N_U$ matrix formed by taking
$\zvec{k}$ as the $k^{\rm th}$ column, i.e.
\begin{eqnarray} 
\boldsymbol{A}_{\rm eq} = \left[ \begin{array}{c|c|c|c}
      & & & \\
     \zvec{0} & \zvec{1} & \cdots & \zvec{N_U} \\
      & & & \\
    \end{array} \right]
\,.
\end{eqnarray}
The solution $\boldsymbol{x}$ to this linear programming problem would give
the weights $w_k$ of the unitaries $U_k$ used to achieve the desired
goal.

However, while this formulation brings our search problem into
standard form for linear programming, this construction does not
include additional terms which would minimize the number of unitaries,
or minimize the total weight employed.  These can usually be included
using other constraint mechanisms available within an optimization
package. Also, the weights $w_k$ are typically desired to be integers.  With
this constraint, the optimization problem becomes a mixed integer
programming problem, which is harder to solve than the relaxed,
linear programming formulation given above.

\subsection{Optimization formulation}

We now introduce an objective function into the formulation.  With the goal to maximize the proportionality constant $\beta$ of the desired Hamiltonian $H_1$, one could choose to minimize $\boldsymbol{A}_{\rm ub} \boldsymbol{x}$.  However, with this requirement encoded in a constraint, we use the objective function for cardinality and weight regularization.  Incorporation of a cardinality penalty term offers control over the overall size of the set of pulse primitives (dictionary) $\mathcal U_{\rm seq}$,
\begin{eqnarray}
\min_{x_i \in \mathbb{Z}^+} & \sum x_i &+ \alpha \|\boldsymbol{x}\|_0 \\
{\textrm {s.t.} } & \boldsymbol{A}_{\rm eq} \boldsymbol{x} &= \boldsymbol{b}_{\rm eq} \nonumber\\ 
& \boldsymbol{A}_{\rm ub} \boldsymbol{x} &\le \boldsymbol{b}_{\rm ub} \nonumber\\
& 0 \le  \ x_i  &\le  u_{i} \nonumber
\end{eqnarray}
where the parameter $\alpha$ may be tuned to weight cardinality versus total sequence length, and $u_i$ is an upper bound on $x_i$.  The $\mathit{l}_0$ quasinorm $\|\boldsymbol{x}\|_0 = \sum_i |x_i|$ is a nonlinear function, so to incorporate such a penalty term in a conventional linear programming (LP) or integer programming (IP) framework, we introduce a set of binaries, $z_i$, satisfying $0 \le x_i \le ub_i \cdot z_i$, where $x_i$ can either be general integer or continuous variable, ${\rm ub}_i$ is the  upper bound of $x_i$. Consequently, a value of $z_i = 0$ implies $x_i = 0$. Thus, to add a cardinality penalty we can simply add a term $\sum_i z_i$ to the objective. 
\begin{eqnarray}
\min_{x_i \in \mathbb{Z}^+, z_i \in \mathbb{B}} & \sum \big ( x_i & + \underbrace{\alpha z_i}_{\textrm{cardinality}} \big ) \\
{\textrm {s.t.} } & \boldsymbol{A}_{\rm eq} \boldsymbol{x} &= \boldsymbol{b}_{\rm eq} \\
& \boldsymbol{A}_{\rm ub} \boldsymbol{x} &\le \boldsymbol{b}_{\rm ub} \\
& 0 \le  \ x_i & \le z_i \cdot u_{i}  
\end{eqnarray}
If the right-hand side assignment of variables is not supported on the designated optimization solver (e.g. Matlab CPLEX version), it is advisable to define an extension of the inequality matrix $\boldsymbol{A}_{\rm ub}$, with diagonal of ones (for $x_i$) and a shifted diagonal (by number of elements in $\boldsymbol{x}$) with $\boldsymbol{b}_{\rm ub}$ as follows
\begin{eqnarray}
\min_{x_i \in \mathbb{Z}^+, z_i \in \mathbb{B}} &  \sum \big ( x_i  + \underbrace{\alpha z_i}_{\textrm{cardinality}} \big ) \\
{\textrm {s.t.} } & 
\left[\begin{array}{c|c}
    \boldsymbol{A}_{\rm eq} & \boldsymbol{0}
\end{array}\right]
\left[\begin{array}{c}
    \boldsymbol{x} \\
    \boldsymbol{z}
\end{array}\right]
  = \boldsymbol{b}_{\rm eq} \\
& 
\left[\begin{array}{c|c}
    \boldsymbol{A}_{\rm ub}  & \mathbf{0} \\    
    \boldsymbol{I}      & -\boldsymbol{I} \odot \boldsymbol{b}_{\rm ub}
\end{array}\right]
\left[\begin{array}{c}
    \boldsymbol{x} \\
    \boldsymbol{z}
\end{array}\right]
 & \le  
\left[\begin{array}{c}
    \boldsymbol{b}_{\rm ub} \\
    \mathbf{0}
\end{array}\right] \\
& \; 0 \le  \ 
\left[\begin{array}{c}
    x_i \\
    z_i
\end{array}\right]
\le \left[\begin{array}{c}
    u_{i}  \\
    1
\end{array}\right] \\
\end{eqnarray} 
where $\odot$ denotes the Hadamard product.

\section{Dipolar decoupling for broadband magnetometry
\label{sec:dip_decoup_for_broad_mag}}

We now consider the problem of decoupling inhomogeneous dipolar interactions in spin ensembles while retaining sensitivity to an external magnetic field.  This section begins with a model of a spins in a solid subject to a global magnetic field, with dipole-dipole interactions between spins, followed by the definition of decoupling pulse sequence success criteria for ensemble magnetometry.  
A spin-1/2 qubit model provides geometrical intuition, and we present a pulse sequence that achieves the success criteria.  Applying the previously developed optimization formulation to the spin-1 qutrit model generates a family of pulse sequences, and we examine one in particular with the most desirable characteristics for broadband magnetometry.  

\subsection{Spin ensemble model
\label{subsec:spin_ensemble}}

The dynamics of a solid-state spin ensemble are governed by an external magnetic field $\mathbf B$ and interactions between spins $J_{ij}$, shown schematically in \fref{fig:ensemble_schematic_FID_spectrum}(a).  The spins in the system are generally subject to an external magnetic field 
\begin{equation}
    \mathbf B = \mathbf B_0 + \mathbf B_{\rm sense}
\end{equation}
that is the sum of a DC bias field $\mathbf B_0$ which is known in principle, and an unknown field of interest $\mathbf B_{\rm sense}$ which the sensor measures, where $B_{\rm sense} \ll B_0$.  The spins are coupled to each other through dipolar interactions.  The Hamiltonian describing the coupled spin system 
\begin{equation}
    H = H_{\rm Z} + H_{\rm dd} 
    \label{eq:H}
\end{equation}
contains a Zeeman term
\begin{equation}
    H_{\rm Z} = \gamma B_z \sum_i S^i_{z}
    \label{eq:H_Z}
\end{equation}
where $B_z$ is the projection of the magnetic field onto the quantization axis $z$, with $\gamma$ the spin gyromagnetic ratio, and a dipolar interaction term
\begin{equation}
    H_{\rm dd} = \sum_{ij} J_{ij} \left ( 3 S^i_{z} S^j_{z} - \mathbf S^i \cdot \mathbf S^j \right ),
    \label{eq:H_dd}
\end{equation}
where $\mathbf S^i = (S^i_{x}, S^i_{y}, S^i_{z})$ is the spin operator for the $i^{\rm th}$ particle of spin $S^i$.  
The dipolar interaction
\begin{equation}
J_{ij} = \frac{J_0}{r_{ij}^3}(1-3\cos^2\theta_{ij})
\end{equation}
depends on the geometrical orientation and physical properties of the two spins, with 
$r_{ij}$ the distance between spins, 
$\theta_{ij}$ the angle between the separation vector and the quantization axis,
and $J_0$ the interaction strength.
  
For noninteracting spins, $J_{ij} = 0$, Ramsey fringes of constant amplitude oscillate sinusoidally in time at a Zeeman frequency $\omega = \gamma B_z$, with the Fourier transform producing a spectral line at this frequency.  Averaging over Gaussians with linewidths $J_{ij}$ drawn from a distribution of coupling strengths 
\begin{equation}
\label{eq:linewidth_distribution}
    P(J) = \frac{\Gamma}{J^2}\sqrt{\frac{2}{\pi}}e^{-\Gamma^2/2J^2}
\end{equation}
representative of an ensemble~\cite{Dobrovitski2008} results in free induction decay of the Ramsey fringes in time and broadening of the spectral line.  
\Fref{fig:ensemble_schematic_FID_spectrum}(c) and (d) compare the free induction decay and corresponding frequency spectrum, respectively, of these limiting cases.  We take $\gamma B_z = 2\pi$, $\Gamma = 2\pi\times10^{-2}$, and average over 10,000 values of pairwise interactions $J_{ij}$ for each case.  Time dependent simulations obtained using QuTiP~\cite{Johansson2012,Johansson2013} calculate the state at time $t = n\,dt$ using the time evolution operator $U(dt)$, given by \eref{eq:U_tc_Hk}, along with the Floquet property, \eref{eq:U_floquet}.

\subsection{Success criteria}
We define the success criteria for dipolar decoupling in spin ensemble magnetometry as follows.  
\begin{itemize}
\item{\bf Clean Zeeman} - the Ramsey signal of an experimental observable is a pure, single frequency sinusoidal oscillation
\item{\bf Zeeman Strength} - the scaling factor $\beta$ of the Ramsey signal frequency of an experimental observable with respect to the same observable under ideal Zeeman evolution
\end{itemize}
These criteria express achievability and optimality conditions defined in \sref{subsec:search_crit}, including \eref{eq:searchgoal}.  The search succeeds when it finds a set of unitaries that transforms the system Hamiltonian, \eref{eq:H}, to one proportional to the desired Zeeman terms $H_{\rm Z}$ only, and averages the undesired dipolar coupling terms $H_{\rm dd}$ to zero, which indeed results in an experimentally observable clean Zeeman signal.  Optimality includes maximizing the scaling factor $\beta$ of the resulting desired terms.  This is precisely the scaling of the Ramsey fringe frequency with respect to a given magnetic field.  

We will find it useful to compare the Zeeman strength of decoupling sequences that achieve the clean Zeeman criterion with existing dipolar decoupling sequences that do not meet the clean Zeeman requirement.  The Zeeman strength for a Hamiltonian that is not a clean Zeeman is calculated by normalizing the projection onto the generalized Pauli basis, and taking the inner product with $S_z$.  

\subsection{Qubit ensembles \label{subsec:qubit_ensembles}}

For spin-1/2 qubits, the spin matrices $S_{k} = \sigma_{k}/2$ are proportional to the Pauli matrices $\sigma_{k}$ for $k = x, y, z$.  Rotation operators about each axis by an angle $\theta$ are 
\begin{equation}
    R_{k}(\theta) = \exp(-i\sigma_k\theta/2)
\end{equation}
and for $\pi/2$ rotations we adopt the following notation
\begin{equation}
    X = R_x(\pi/2), \quad \bar{X} = R_x(-\pi/2),
\end{equation}
analogously defined for $y$ and $z$.  The set of rotations that maps Pauli operators to Pauli operators (up to a multiplicative factor $\pm 1$) forms a group, known as the Clifford group.  Let us formalize this set by defining the operators,
\begin{equation}
\def\arraystretch{1.25}
\hspace{-5.5pt}
\begin{array}{l}
    W_0 = I,\ W_1 = X,\ W_2 = X^2,\ W_3 = \bar X, \\
    V_0 = I,\ V_1 = Z,\ V_2 = \bar Y,\ V_3 = Z^2,\ V_4 = \bar Z,\ V_5 = Y,
    \label{eq:clifford_generators}
\end{array}
\end{equation}
from which we may construct any of the 24 Clifford rotations from the product $V_i W_j$.  Thinking of the Clifford group as the set of operations that maps the $x,y,z$ axes, or corresponding Pauli operators, to distinct orientations along the cardinal axes $\pm x, \pm y, \pm z$ axes provides geometrical intuition.  Consider, for example, that the six $V_i$ map $\sigma_x$ to unique cardinal directions, and the four $W_j$ orient $\sigma_y$ and $\sigma_z$ with respect to $\sigma_x$.  

\begin{figure*}
\includegraphics[width=\textwidth, trim={0 2.5cm 0 1cm},clip]{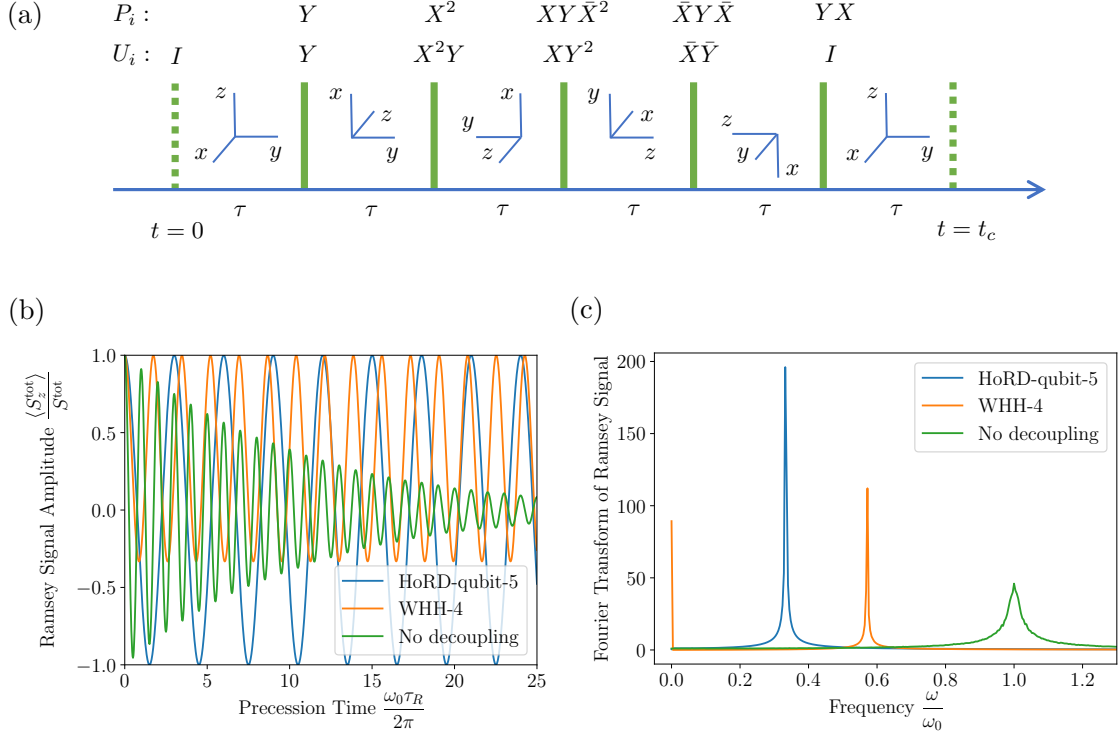}
\caption{\textbf{HoRD-qubit-5 pulse sequence.} (a) Pulses $P_i$ and corresponding unitary basis transformation operators $U_i$ for the HoRD-qubit-5 pulse sequence.  The $x, y, z$ axes show the toggling frame transformation of the corresponding Pauli operators $\sigma_{x,y,z}$ during each interval. (b) Free induction signal and (c) corresponding spectrum of a Ramsey experiment where the HoRD-qubit-5 pulse sequence is applied during the free evolution time, with WHH-4~\cite{Waugh1968} and no decoupling shown for comparison.  The HoRD-qubit-5 pulse sequence decouples dipolar interactions in spin-1/2 ensembles, resulting in a constant amplitude single frequency sinusoid (clean Zeeman) with frequency scaled by a factor of $\beta = 1/3$ (Zeeman strength) relative to $\omega/\omega_0 = 1$, where $\omega_0 = \gamma B_z$.  By comparison, the WHH-4 sequence does not produce clean Zeeman behavior, as it reduces the amplitude of the sinusoid and produces a DC offset, resulting in a spectral component at $\omega = 0$ in addition a component scaled by a factor of $\beta = 1/\sqrt{3}$.  
\label{fig:qubit_wahuha_d3cm_FID_spectrum}}
\end{figure*}

The WHH-4 sequence~\cite{Waugh1968} 
\begin{equation}
    \tau \bar{X} \tau Y 2\tau \bar{Y} \tau X \tau
\end{equation}
known from NMR decouples homonuclear dipolar interactions to leading order.  The instantaneous pulses punctuate free evolution time intervals $\tau_i = \tau$ for $i = 0, 1, 3, 4$ and $\tau_2 = 2\tau$.  The time ordering of the free evolution times and pulses goes from right to left.  The corresponding unitary basis transformations ($U_k = P_k \cdots P_0$)
\begin{equation}
    U_4 = I,\ U_3 = X,\ U_2 = \bar{Y}X,\ U_1 = X,\ U_0 = I,
\end{equation}
are elements of the Clifford group, $V_i W_j$ as defined in \eref{eq:clifford_generators}. 
Substituting these unitary operators in the leading order average Hamiltonian term, \eref{eq:avg_hamiltonian_0}, applied to the dipole-dipole interaction Hamiltonian, \eref{eq:H_dd}, gives $\bar{H}^{(0)}_{\rm dd} = 0$, \eref{eq:avg_hamiltonian_1}.  However, WHH-4 also depolarizes the Zeeman term, \eref{eq:H_Z}, giving $\bar{H}^{(0)}_{\rm Z} = (S_x + S_y + S_z)/3$, isotropically projecting it equally onto all three spin components and reducing its magnitude.  
Applying the WHH-4 sequence during a Ramsey experiment produces a signal
$\langle S_z^{\rm tot}\rangle/S_z^{\rm tot} = \frac{1}{3} + \frac{2}{3}\cos(\omega_0 \tau_R/\sqrt{3})$, shown in \fref{fig:qubit_wahuha_d3cm_FID_spectrum}.  
Note the zero frequency component in addition to Ramsey oscillations scaled in amplitude by a factor of $2/3$, and in frequency by $1/\sqrt{3}$.  The frequency scaling is consistent with calculation of the Zeeman strength $\beta = 1/\sqrt{3}$ obtained by normalizing the projection of the Hamiltonian onto the Pauli matrices and taking the inner product with $S_z = \sigma_z/2$.  

While WHH-4 works in practice for homonuclear decoupling in NMR, given that low frequencies are effectively filtered out, for magnetometry, it may prove suboptimal.  
Applying a decoupling pulse sequence that effectively decouples dephasing interactions increases the available measurement time $\tau$, but also alters the effective gyromagnetic ratio $\gamma$ and measurement contrast $C$ must also be accounted for.  
The magnetic field sensitivity, \eref{eq:sensitivity} is inversely proportional to the maximum slope $\gamma C$ of the Ramsey signal.  
For WHH-4, $\gamma C = \frac{2}{3\sqrt{3}} \approx 0.385$, where $\gamma$ and $C$ capture the scaling of the gyromagnetic ratio and contrast, respectively, due to the pulse sequence.  
In cases where the contrast noise is independent of the signal amplitude, e.g. photon shot noise limited optically detected magnetic resonance, reducing the contrast is generally undesirable.  
The zero frequency component may limit the use of WHH-4 in low or zero bias field operation, of current interest for certain applications~\cite{Backlund2017, Munzhuber2017}.  

The following dipolar decoupling for broadband magnetometry sequence, which we call HoRD-qubit-5,
\begin{equation}
    \tau YX \tau \bar{X}Y\bar{X} \tau XY\bar{X}^2 \tau X^2 \tau Y \tau,
\end{equation}
found analytically using average Hamiltonian theory, yields a single frequency Ramsey oscillation (\fref{fig:qubit_wahuha_d3cm_FID_spectrum}).  The corresponding basis transformations
\begin{equation}
    U_5 = I,\ U_4 = \bar{X}\bar{Y},\ U_3 = XY^2,\ U_2 = X^2 Y,\ U_1 = Y,\ U_0 = I
\end{equation}
with free evolution for intervals $\tau_i = \tau$ for all $i$ averages $H_{\rm dd}$ to zero while scaling the strength of $H_{\rm Z}$ by a factor of $1/3$.  With no change in contrast, this gives a maximal slope $\gamma C = \frac{1}{3} \approx 0.333$ slightly lower yet comparable to WHH-4.  The HoRD-qubit-5 pulse sequence also zeros the second order average Hamiltonian, \eref{eq:avg_hamiltonian_1}.  These unitary operators are in fact Clifford operators, $V_i W_j$, defined in \eref{eq:clifford_generators}, where relations such as $R_z(\theta) = \bar X R_y(-\theta) X = \bar Y R_x(\theta) Y$ are useful in showing their equivalence, e.g. $U_4 = \bar X \bar Y = Z \bar X = V_1 W_3$.  \Fref{fig:qubit_wahuha_d3cm_FID_spectrum}(a) depicts the orientation of the Pauli operator axes during each interval, taking advantage of the mapping of $SU(2)$ rotations in spin space to $SO(3)$ rotations in real space. Note that any sequence that orients the $+z$ axis along $\pm x$, $\pm y$, $\pm z$, for equal intervals achieves the desired leading order decoupling for a static relative arrangement of dipoles. This freedom to tailor the pulse sequence may prove useful in canceling higher order terms or simplifying the experimental control pulses.  Time-dependent simulations show that HoRD-qubit-5 results in a single frequency Ramsey signal $\langle S_z^{\rm tot}\rangle/S_z^{\rm tot} = \cos(\omega_0 \tau_R/3)$ and corresponding spectrum with a single resonance at $\omega/\omega_0 = 1/3$ (\fref{fig:qubit_wahuha_d3cm_FID_spectrum}(b) and (c), respectively).   

The HoRD-qubit-5 sequence would find application in magnetometry with dense spin-1/2 ensembles for which dipolar coupling between spins is a limiting factor for phase coherence.  The negatively charged silicon vacancy center (SiV$^{\text -}$) in diamond is a spin-1/2 color center~\cite{Hepp2014}, for which coherence of a single SiV$^{\text -}$ is limited by coupling to a spin bath of substitutional nitrogen atoms at sufficiently low (millikelvin) temperatures~\cite{Becker2018}, and by phonon-mediated interactions at higher temperatures~\cite{Becker2017}.  Note that while the qubit approximation to higher spins, in which only a pair of levels is considered, can describe the Zeeman splitting (with the appropriate gyromagnetic ratio), it fails to capture physics of dipole-dipole coupling.  

\subsection{NV ensemble magnetometry}

Demonstrations of sensitive broadband magnetometry~\cite{Barry2016,Glenn2018}, and recent progress elucidating and eliminating dominant dephasing mechanisms~\cite{Bauch2018} in these systems, make dense NV spin ensembles particularly interesting for applying the dipolar decoupling techniques investigated in this work.  Here we describe some essential physics of NV ensembles relevant to magnetometry, discuss how these systems relate to the spin ensemble model in \sref{subsec:spin_ensemble}, delve into the utility of the spin-1 property of the NV$^{\text -}$, and consider the classical control available in state of the art experiments.  

\begin{figure*}
\includegraphics[width=\textwidth, trim={0 2.5cm 0 1cm},clip]{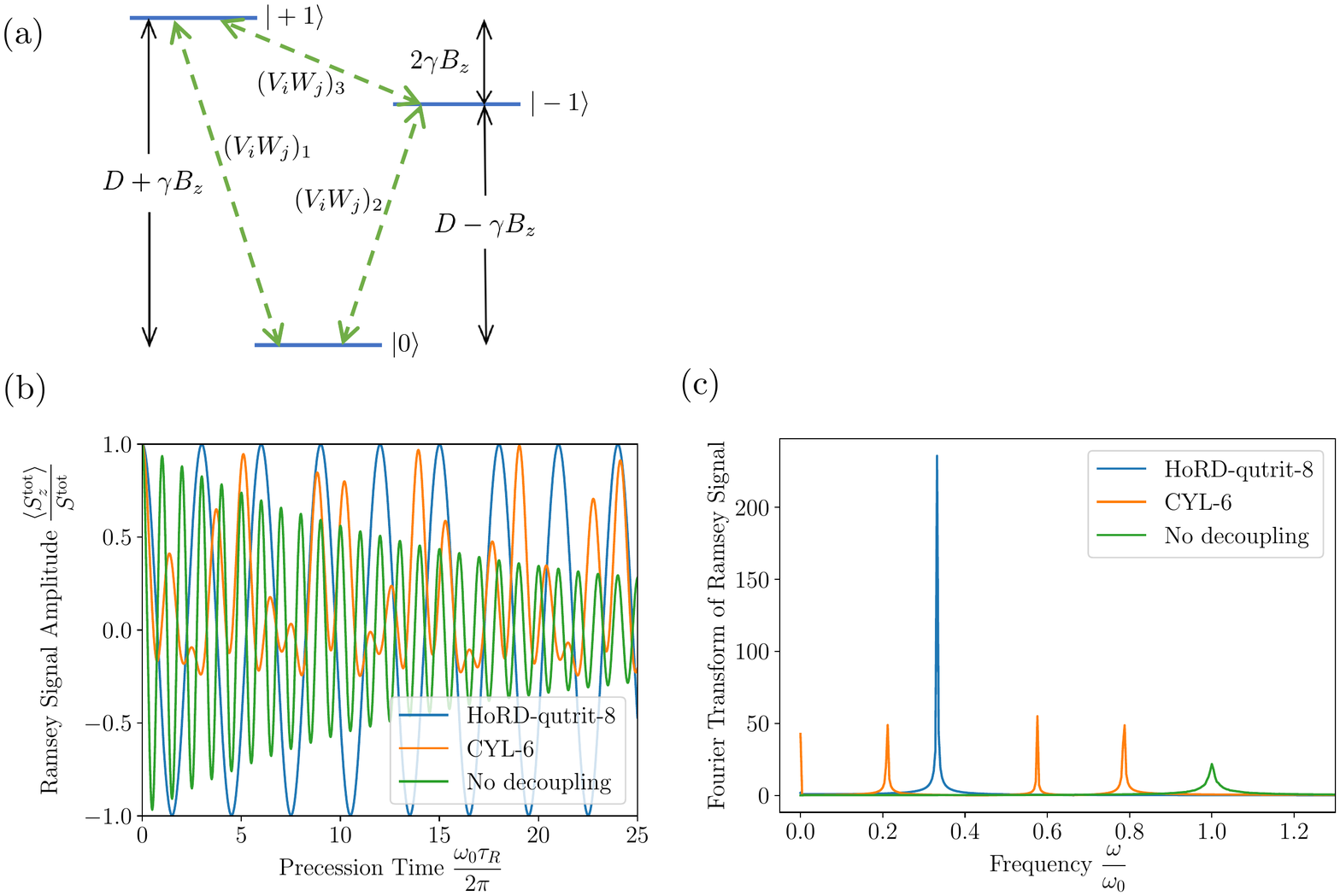}
\includegraphics[width=\textwidth, trim={0 2.5cm 0 8.25cm},clip]{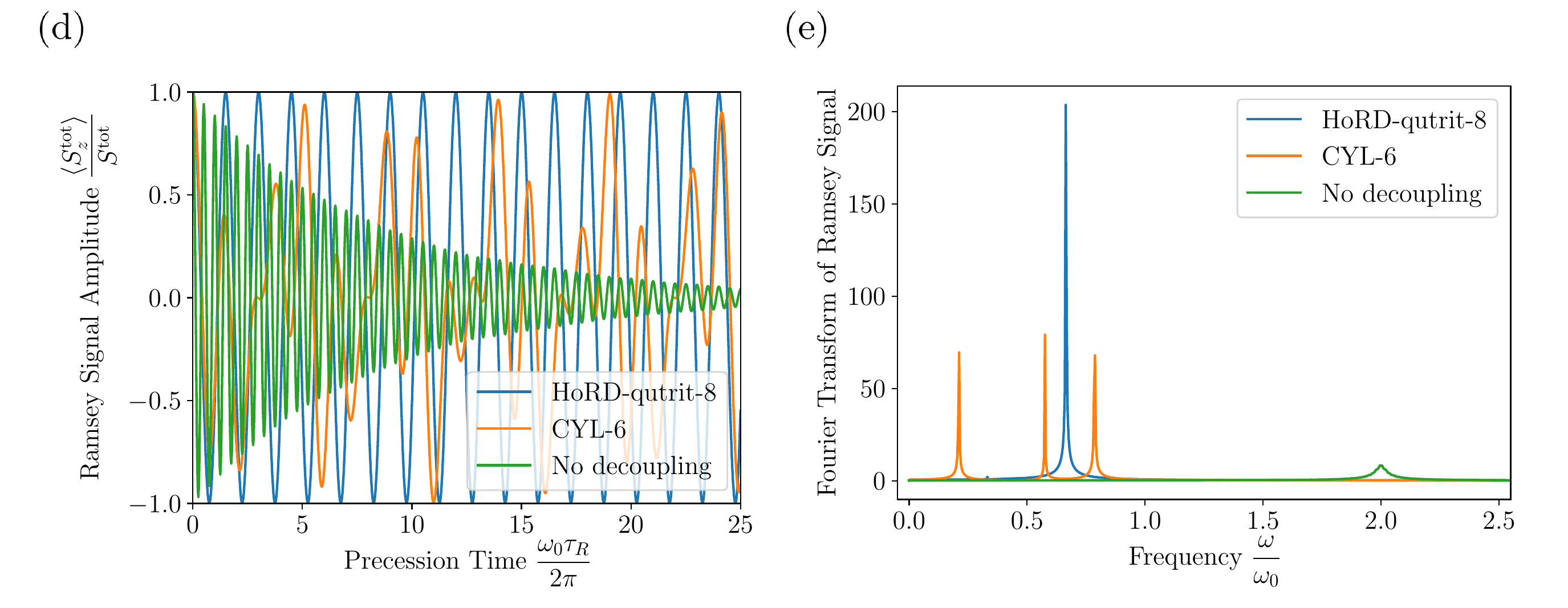}
\caption{\textbf{HoRD-qutrit-8 pulse sequence.} (a) NV energy level diagram with two-level Clifford operations $(V_iW_j)_k$ between the three  subsystems $k = 1,2,3$. (b), (d) Free induction signal and (c), (e) corresponding spectrum of HoRD-qutrit-8 in the single quantum (SQ, $|\psi_0\rangle = (|0\rangle+|1\rangle)/\sqrt{2}$) and double quantum (DQ, $|\psi_0\rangle = (|1\rangle+|-1\rangle)/\sqrt{2}$) bases, respectively, of a Ramsey experiment with the HoRD-qutrit-8 pulse sequence applied during the free evolution time, with CYL-6~\cite{Choi2017} and no decoupling shown for comparison.  The HoRD-qutrit-8 pulse sequence decouples dipolar interaction in a spin-1 ensemble and produces a clean Zeeman signal with frequency scaled by a factor of the Zeeman strength $\beta = 1/3$ relative to $\omega/\omega_0 = 1$ ($\omega/\omega_0 = 2$) for SQ (DQ) magnetometry, where $\omega_0 = \gamma B_z$.  The CYL-6 sequence also performs dipolar decoupling, but produces a Ramsey signal containing multiple frequency components which also depend on the choice of basis.  Note that without decoupling, inhomogeneous interactions produce a faster decay and broader linewidth for DQ compared to SQ.  
    \label{fig:nv_cyl_sr10zeeman_FID_spectrum}}
\end{figure*}

The NV$^{\text -}$ ground state manifold Hamiltonian $H_{\rm NV} = H_{\rm Z} + H_D$ contains the Zeeman term $H_{\rm Z} = \gamma B_z S_z$ and the zero field splitting $H_D = 2\pi D S_z^2$.  $B_z$ is the component of the magnetic field projected onto the NV$^\text{-}$ axis, $D \approx 2.87$ GHz is the zero-field splitting, and $S_z$ is a spin-1 operator.  \Fref{fig:nv_cyl_sr10zeeman_FID_spectrum}(a) shows the NV energy levels.  In an ensemble containing all four NV orientations, the magnetic field projects differently in principle onto each axis.  The model considered here corresponds to the magnetic field oriented along a single NV$^\text{-}$ class.  
This can be realized by optically initializing all NV$^{\text -}$ orientations to the $|0\rangle$ state, then preparing a superposition of only a single chosen orientation.

Tailoring the growth process to achieve preferential NV orientation along two axes~\cite{Edmonds2012} or a single axis~\cite{Fukui2014,Michl2014,Lesik2014,Tahara2015} has resulted in increased measurement contrast in dilute~\cite{Pham2012a} and dense~\cite{Ozawa2017} ensembles, yet the irradiation and annealing used to optimize NV$^\text{-}$ concentration destroys preferential orientation.  Note that hyperfine interactions introduce beats in a typical NV$^\text{-}$ Ramsey signal; the current analysis considers only the electronic spin degrees of freedom.  Other color centers to which the current result applies include neutrally charged silicon vacancies~\cite{Iakoubovskii2002,Edmonds2008} in diamond and divacancies in silicon carbide~\cite{Heremans2016}.   

To connect the NV model to the spin ensemble model, \eref{eq:H}, we analyze the effect of the zero field splitting.  Moving to the rotating frame of $H_D$ leaves $H_Z$ unchanged, and shows that the large zero-field splitting of the NV suppresses certain dipole-dipole matrix elements.  This makes any sequence that decouples dipolar interactions in a qutrit ensemble model applicable to NV$^\text{-}$ ensembles, whereas the converse is not necessarily true.  

The spin-1 property of the NV$^\text{-}$ provides three choices of superposition state for the Ramsey measurement, enabling both single quantum (SQ) and double quantum (DQ) magnetometry.  SQ magnetometry uses a superposition of $|0\rangle$ and $|+1\rangle$, or $|0\rangle$ and $|-1\rangle$ states.  Because the zero-field splitting $D$ has temperature dependence and is broadened by inhomogeneous strain, this basis is suboptimal for broadband magnetometry.  DQ magnetometry employs a superposition of $|+1\rangle$ and $|-1\rangle$ states in order to cancel out common-mode shifts due to temperature fluctuations or strain inhomogeneity.  The doubled effective gyromagnetic ratio results in a Ramsey frequency that exhibits a two-fold increase, and dephasing due to dipolar broadening increases by a factor of two for spin bath coupling and four for NV$^\text{-}$-NV$^\text{-}$ coupling, which can be seen in \fref{fig:nv_cyl_sr10zeeman_FID_spectrum}~(b)-(e).  This amplifies the requirement for a pulse sequence to produce an average Hamiltonian containing the full $S^{\rm tot}_z = \sum_i S^i_z$ Zeeman term, in order to produce a clean Zeeman Ramsey signal for any of the three superposition state bases.  

The quantum states of a given class of NV$^\text{-}$ in an ensemble can be controlled by delivering resonant microwave pulses to the entire sample.  Transitions between 0 and $\pm$1 are driven directly at frequencies $D \pm \gamma B_z$, whereas transitions between $\pm1$ states require composite microwave pulses.  Current experimental setups include multiple microwave drive channels, each with independent phase and amplitude control, and well-tuned $\pi/2$ and $\pi$ pulses suitable for SQ and DQ magnetometry~\cite{Neumann2013,Bauch2018}.

\subsection{Qutrit ensembles
\label{subsec:qutrit}}
With the physics of NV systems, relationship to the qutrit ensemble model, and available controls established, we now develop an explicit model for the spin-1 quantum pulse sequence search problem.  The increased dimensionality of spin-1 qutrit ensembles both affords greater flexibility for Hamiltonian engineering and requires a more systematic approach to push beyond the geometrical intuition apparent for spin-1/2 systems. 

Consider Clifford rotations between pairs of levels $\{+1,0\}$, $\{0,-1\}$, $\{+1,-1\}$ with $k = 1,\ 2,\ 3$ corresponding to each qubit subsystem, respectively,  
The Gell-Mann matrices $\lambda_k$, $k = 1, \ldots, 8$, given explicitly in \nameref{sec:appendix_A}, provide an orthonormal basis for traceless Hermitian $3\times3$ matrices, and are the generators of $SU(3)$.  
The rotation operators
\begin{eqnarray}
    (R_x(\theta))_k =  \exp(-i\lambda_k\theta/2), \quad k = 1,2,3, \\
    (R_y(\theta))_k =  \exp(-i\lambda_{k+3}\theta/2), \quad k = 1,2,3, \\
    (R_z(\theta))_1 =  \exp(-i\lambda_7\theta/2), \\
    (R_z(\theta))_2 =  \exp(-i\zeta\theta/2), \\
    (R_z(\theta))_3 =  \exp(-i\eta\theta/2),
\end{eqnarray}
for each subsystem are built from the corresponding $SU(2)$ subalgebra, \eref{eq:su2_subalgebra}, where $\zeta$ and $\eta$ are linear combinations of $\lambda_7$ and $\lambda_8$.  The Clifford group for subsystem $k$ is defined by using the rotation operators 
\begin{equation}
    (X)_k = (R_x(\pi/2))_k, \quad (\bar{X}) = (R_x(-\pi/2))_k,
\end{equation}
analogously defined for $y$ and $z$.
in the definitions of $\{V_i, W_j\}$ in \eref{eq:clifford_generators}, with the substitution $V_i W_j \rightarrow (V_i W_j)_k$.  While we consider only Clifford rotations here given the prevalence of pulsed quantum control of NV ensembles~\cite{Bauch2018}, this restriction is not fundamental.  Quantum control techniques that are robust against noise, such as fast holonomic gates~\cite{Sjoqvist2012}, which have recently been demonstrated in NV systems~\cite{Arroyo-Camejo2014}, present an interesting avenue for future work.  

With a clean Zeeman target Hamiltonian, we search for and find a set of unitary operators.  We harness the IBM ILOG CPLEX~\cite{CPLEX} library that handles linear programming (LP), integer programming (IP), and constraint programming (CP).  Unitary operators made from a product of three operators between sublevels $U = (V_i W_j)_1 (V_k W_l)_2 (V_m W_n)_3$ comprise the search space for each qutrit unitary, giving $24^3 = 13,824$ possibilities for each unitary in the sequence.  Tabulating how every unitary operator transforms the Hamiltonian $H$ given by \eref{eq:H} as $U^\dagger H U$ shows 558 unique mappings, allowing us to significantly prune the search set.  We anticipate sequences containing at least six unitaries, such as CYL-6, needed to decouple dipolar interactions and possibly longer sequences to obtain a clean Zeeman term.

We discover a family of sequences for qutrit ensembles that decouples dipolar interactions with various resulting Zeeman terms, which we describe here.  Linear programming finds the set of twelve unitary operators, given explicitly in \nameref{sec:appendix_B}.  While this set does zero dipolar interactions, it also undesirably zeros the Zeeman term exactly ($\beta = 0$), so we call it Homonuclear zero Zeeman Decoupling (HoZD-qutrit-12).  
However, analyzing how each unitary $U_k$ transforms $H_{\rm Z}$ and $H_{\rm dd}$ to leading order as $\bar H^{(0)} = \sum_k U_k^\dagger H U_k$ shows that it can lead to a clean Zeeman Hamiltonian.  Each pair of terms $U_i$, $U_{i+1}$ for even $i$ has the same action on the dipolar coupling, and maps the Zeeman term to the same Gell-Mann matrix but with opposite sign, $U_i^\dagger S_z U_i = (-1)^i \lambda_j$, $j = \lfloor\frac{i}{2}\rfloor+1$.  Thus, we can choose which Zeeman terms to turn on and off without sacrificing dipolar decoupling.  For example, $U_0^\dagger S_z U_0 = \lambda_1$ and $U_1^\dagger S_z U_1 = -\lambda_1$.  Removing $U_1$ and doubling the interval after, or equivalently the corresponding weight of, $U_0$ gives $\bar H^{(0)} = \frac{1}{6} \gamma B_z \lambda_1$.  Because the desired Hamiltonian should be proportional to $S_z$, we search for a unitary rotation from the same set of sub-level Clifford operator combinations and find $U = (V_5 W_0)_1 (V_0 W_2)_3$ gives $U^\dagger \lambda_1 U = \frac{1}{2}(\lambda_7 + \sqrt{3}\lambda_8) = S_z$.  This is a clean Zeeman with Zeeman strength $\frac{1}{6}$.  

To maximize the Zeeman strength, we systematically turn on terms to which the Zeeman Hamiltonian maps, then look for a unitary operator from the search space that maps the average Hamiltonian back to $S_z$.  The Zeeman strength resulting from a set of unitary operators should increase with fewer intervals canceling each other out, so we want to maximize the number of Zeeman interval pairs that are on.  

We find that the sequence $\{U_0, U_2, U_4, U_5, U_6, U_8, U_{10}, U_{11}\}$ with weights $\{2, 2, 1, 1, 2, 2, 1, 1\}$ produces an average Hamiltonian with terms proportional to $\lambda_1 + \lambda_2 + \lambda_4 + \lambda_5$.  A brute force search of products of two-level Cliffords that maps this to $S_z$ finds $U = (V_0W_2)_1 (V_4W_1)_2 (V_1W_1)_3$.  Multiplying every unitary in the sequence by this gives explicitly the result 
\begin{equation}
\def\arraystretch{1.25}
\hspace{-5.5pt}
\begin{array}{l}
    U_0 = (V_4W_2)_1 (V_2W_2)_2 (V_3W_2)_3 (V_0W_2)_1 (V_4W_1)_2 (V_1W_1)_3, \\
    U_1 = (V_4W_2)_1 (V_1W_1)_2 (V_3W_0)_3 (V_0W_2)_1 (V_4W_1)_2 (V_1W_1)_3, \\
    U_2 = (V_1W_0)_1 (V_3W_0)_2 (V_2W_1)_3 (V_0W_2)_1 (V_4W_1)_2 (V_1W_1)_3, \\
    U_3 = (V_0W_0)_1 (V_4W_0)_2 (V_5W_3)_3 (V_0W_2)_1 (V_4W_1)_2 (V_1W_1)_3, \\
    U_4 = (V_0W_2)_1 (V_5W_0)_2 (V_0W_2)_3 (V_0W_2)_1 (V_4W_1)_2 (V_1W_1)_3, \\
    U_5 = (V_1W_2)_1 (V_3W_1)_2 (V_0W_0)_3 (V_0W_2)_1 (V_4W_1)_2 (V_1W_1)_3, \\
    U_6 = (V_3W_0)_1 (V_4W_0)_2 (V_3W_1)_3 (V_0W_2)_1 (V_4W_1)_2 (V_1W_1)_3, \\
    U_7 = (V_4W_0)_1 (V_3W_0)_2 (V_4W_3)_3 (V_0W_2)_1 (V_4W_1)_2 (V_1W_1)_3, 
\end{array}
\end{equation}
with weights $\{w_k\} = \{2, 2, 1, 1, 2, 2, 1, 1\}$.  This set of unitary rotations achieves the Clean Zeeman criteria, $\bar H^{(0)} = \frac{1}{3} \gamma B_z S_z$, with Zeeman strength $\beta = 1/3$.  We call this sequence HoRD-qutrit-8 and it is one of the main results of this work.  The corresponding pulses may be found using $P_k = U_k U_{k-1}^\dagger$,  
\begin{equation}
\def\arraystretch{1.25}
\hspace{-5.5pt}
\begin{array}{l}
    P_0 = (V_4W_2)_1 (V_2W_2)_2 (V_3W_2)_3 (V_0W_2)_1 (V_4W_1)_2 (V_1W_1)_3 \\
    P_1 = (V_4W_2)_1 (V_1W_1)_2 (V_3W_0)_3 (V_3W_2)_3^\dagger (V_2W_2)_2^\dagger (V_4W_2)_1^\dagger, \\
    P_2 = (V_1W_0)_1 (V_3W_0)_2 (V_2W_1)_3 (V_3W_0)_3^\dagger (V_1W_1)_2^\dagger (V_4W_2)_1^\dagger, \\
    P_3 = (V_0W_0)_1 (V_4W_0)_2 (V_5W_3)_3 (V_2W_1)_3^\dagger (V_3W_0)_2^\dagger (V_1W_0)_1^\dagger, \\
    P_4 = (V_0W_2)_1 (V_5W_0)_2 (V_0W_2)_3 (V_5W_3)_3^\dagger (V_4W_0)_2^\dagger (V_0W_0)_1^\dagger, \\
    P_5 = (V_1W_2)_1 (V_3W_1)_2 (V_0W_0)_3 (V_0W_2)_3^\dagger (V_5W_0)_2^\dagger (V_0W_2)_1^\dagger, \\
    P_6 = (V_3W_0)_1 (V_4W_0)_2 (V_3W_1)_3 (V_0W_0)_3^\dagger (V_3W_1)_2^\dagger (V_1W_2)_1^\dagger, \\
    P_7 = (V_4W_0)_1 (V_3W_0)_2 (V_4W_3)_3 (V_3W_1)_3^\dagger (V_4W_0)_2^\dagger (V_3W_0)_1^\dagger, 
\end{array}
\end{equation}
with further simplification possible by reordering the unitary operators and compressing the pulses~\cite{Choi2017}.  Note that one additional pulse $P_8 = U_7^\dagger$ following the final free evolution interval returns the system to its initial state ($P_8 P_7 \cdots P_1 P_0 = I$), and may be combined with the second Ramsey pulse prior to readout.  

With the HoRD-qutrit-8 sequence producing a Hamiltonian having a clean spin-1 Zeeman Hamiltonian form, we may choose any superposition state for the Ramsey measurement, enabling both single and double quantum magnetometry.  \Fref{fig:nv_cyl_sr10zeeman_FID_spectrum} shows the free induction signal and corresponding spectrum of HoRD-qutrit-8 in the single quantum and double quantum bases.  

Analyzing the effect of CYL-6 shows the average Hamiltonian maps as $\bar H^{(0)} = \frac{1}{6}(-\lambda_1 + \lambda_2 - \lambda_4 + \lambda_5 + \lambda_6 + \frac{1}{2}(\lambda_7 + \sqrt{3}\lambda_8))$, which does not produce a clean Zeeman Hamiltonian.  The Zeeman strength of the resulting Hamiltonian is calculated by normalizing the projection onto the Gell-Mann matrices, and taking the inner product with $S_z = \frac{1}{2}(\lambda_7 + \sqrt{3}\lambda_8)$, giving $\beta = \frac{1}{\sqrt{6}}$.  The Zeeman strength $\beta = 1/ \sqrt{6} \approx 0.408$ of CYL-6 is slightly larger than $\beta = 1/3 \approx 0.333$ of HoRD-qutrit-8.  However, resonances of the observable Ramsey signal of CYL-6 depend on the choice of initial state (see \fref{fig:nv_cyl_sr10zeeman_FID_spectrum} for a comparison of SQ and DQ initial state), unlike those of HoRD-qutrit-8, hindering direct comparison of the Zeeman strength.  
 
An optimal pulse sequence will achieve the largest Zeeman strength $\beta \leq 1$ with the fewest number of pulses $|\mathcal U_{\rm seq}|$ and smallest total weight $\sum_k w_k$.  
Proving optimality is difficult in practice.  
Brute force search can place a lower bound on sequence length, but becomes computationally expensive after a few pulses due to combinatorial explosion, as discussed in Section~\ref{sec:ps_search_as_constr_opt}.  
Analyzing known sequences that decouple dipolar interactions but do not produce a clean Zeeman term provides some insight.  
The CYL-6 sequence, and a second six pulse sequence which can be constructed from HoZD-qutrit-12, \eref{eq:qutrit-dd-12}, by taking only even (or odd) terms, $\{U_0, U_2, U_4, U_6, U_8, U_{10}\}$ all with equal weight, are the shortest currently known.  
As neither of these sequences produces a clean Zeeman term, sequence length optimality of the eight pulse HoRD-qutrit-8 seems plausible, if not provable.

\section{Discussion
\label{sec:discussion}}

We now discuss experimental realization of the HoRD-qutrit-8 pulse sequence in NV ensembles, opportunities for future work on spin ensemble sensing, and potential application of the current Hamiltonian engineering approach to address crosstalk in superconducting qubit devices for quantum computation.  

\subsection{Experimental realization in NV ensembles}

A successful demonstration of NV$^\text{-}$-NV$^\text{-}$ decoupling using HoRD-qutrit-8 in NV ensembles with sufficiently high nitrogen density must address the following challenges; similar considerations apply to other solid state color centers such as the silicon vacancy center in diamond.  Prolonging the inhomogeneous dephasing time $T_2^*$ in experiments requires a ``shoot the alligator closest to the boat" approach, which entails mitigating the most dominant mechanism first, then systematically addressing each next-dominant factor.  Consider the contributions of independent dephasing mechanisms to $T_2^*$,
\begin{eqnarray}
\frac{1}{T_2^*} &\simeq \frac{1}{T_2^* \{\text{NV$^\text{-}$-P1}\}} 
    + \frac{1}{T_2^* \{\text{strain}\}}
    + \frac{1}{T_2^* \{\text{temperature}\}} \nonumber \\
    &\quad+ \frac{1}{T_2^* \{\text{NV$^\text{-}$-NV$^\text{-}$}\}} 
    + \frac{1}{T_2^* \{\text{NV$^\text{-}$-NV$^\text{0}$}\}} 
    + \ldots
\end{eqnarray}
where $T_2^* \{\cdot\}$ denotes the dephasing time limit due to a specific mechanism~\cite{Bauch2018}.  The P1 spin bath, strain inhomogeneity, and temperature fluctuations likely dominate NV$^\text{-}$ dephasing in high nitrogen density ensembles.  Successively weaker interactions likely include NV$^\text{-}$-NV$^\text{-}$, NV$^\text{-}$-NV$^\text{0}$, and others which are less well understood.  
P1 driving~\cite{Knowles2013} to eliminate dephasing due to the substitutional nitrogen spin bath, and double quantum magnetometry~\cite{Fang2013,Mamin2014} to eliminate broadening due to strain inhomogeneity and temperature fluctuations, should be employed together~\cite{Bauch2018} with the HoRD-qutrit-8 dipolar decoupling protocol for magnetometry.  
State preparation and manipulation using all three spin-1 basis states has been achieved for Ramsey magnetometry using multi-frequency pulses~\cite{Bauch2018}, and for nanoscale temperature sensing using composite pulses between the 0 and $\pm$1 states to effect a pulse between the $-1$ and $+1$ states~\cite{Neumann2013}.  
While the HoRD-qutrit-8 sequence requires a broader library of pulses compared to these state of the art results, leveraging the same fundamental methods that employ amplitude, phase, and frequency control provides a feasible pathway to implementation.  

Control of the NV$^{\text -}$ spin-1 ground state manifold can be accomplished with microwave driving.  
The optimal time for a Ramsey magnetometry measurement is on the order of the inhomogeneous spin dephasing time $T_2^*$.  
P1 spin bath driving combined with DQ magnetometry has resulted in improving native $T_2^* \approx 2 \mu s$ to $T_{2\text{, DQ+Drive}} \approx 30 \mu s$~\cite{Bauch2018}, which we use to estimate the power required for fast pulses.  
A Rabi frequency of $\Omega_R = 2\pi\times 7.7 \text{ MHz}$ has been obtained using a loop gap resonator with incident microwave power $P \approx 16\text{ W}$~\cite{Eisenach2018}, with $\Omega_R \geq 10$ MHz commonly achieved using a wire loop antenna applying a homogeneous field over the sample detection volume~\cite{Pham2012a}.  
Consider the HoRD-qutrit-8 sequence applied once during a total measurement time $\tau_R = 32$ $\mu$s, with composite pulses consisting of six 50 ns-long pulses punctuating free evolution intervals $\tau \approx 4 \mu$s.  This makes the pulses short compared to the free evolution intervals, and the timescale for one cycle $t_c \simeq \tau_R$ short compared to the characteristic dipolar interaction strength $J_0 = 2\pi\times 16\text { kHz/ppm}$, even at nitrogen concentrations exceeding 10 ppm.  
Using multi-tone driving~\cite{Bauch2018} and compressing the pulses, a subject of ongoing investigation, could shorten each composite pulse to 50 ns, making it feasible to repeat the pulse sequence multiple times during the same total measurement time while limiting the accumulated pulse duration to 10\% of the free evolution time.  
For comparison, dynamical decoupling sequences with eight pulses have been performed in less than 2 $\mu$s, and sequences of up to 256 pulses have been demonstrated to increase the coherence time $T_2$ to millisecond time scales, approaching the spin-lattice relaxation $T_1$ limit~\cite{Pham2012}.  

\subsection{Future work}
Avenues of further inquiry for spin ensemble sensing include spin bath decoupling, the effect of multiple sensing spin orientations, and interaction-enhanced metrology.  
Simultaneously driving multiple bath spin and sensing spin resonances does begin to crowd the frequency space and require additional control electronics.  
An additional direction of research will focus on spin bath decoupling for broadband magnetometry using only NV and bias field control, without directly driving the spin bath. 
While the  model considered here describes a single NV orientation or a preferentially oriented NV ensemble with a co-oriented bias field, many ensembles contain NVs equally distributed among the four crystal axes, a critical characteristic for vector magnetometry.  
The current constrained optimization approach will readily incorporate a model with all four NV orientations, yet the existence and, if found, practical utility of such a sequence both remain open questions.  
Interactions play a key role in quantum-enhanced measurements beyond the standard quantum limit, which generally require highly non-classical states with entanglement or squeezing~\cite{Giovannetti2004,Choi2018}.  
Engineering rather than eliminating interactions represents a straightforward extension of the current formalism, although high fidelity entangled state preparation in large systems remains an experimental challenge.  
Ancilla-assisted frequency upconversion presents an alternative scheme for DC magnetometry~\cite{Ajoy2016}.  

To illustrate the generality of this approach, we briefly discuss a second example that arises in superconducting qubit circuits with multiple qubits coupled to the same resonator~\cite{McKay2016}.  Optimizing the fidelity of resonator-mediated two qubit gates between a given pair of qubits requires maximal effective coupling between this pair while eliminating, or reducing to a negligible level, crosstalk with other qubits~\cite{Sheldon2016}.  Definitions of success for the qubit case such as ``exclusive coupling'' and ``coupling strength'' yield appropriate optimization criteria, with fidelity under realistic conditions providing an ultimate performance metric.  

\section{Conclusion
\label{sec:conclusion}}
We cast Hamiltonian engineering as a quantum pulse sequence search problem, identifying desired and undesired terms within the system Hamiltonian, and using average Hamiltonian theory to define achievability and optimality conditions.  Defining orthonormal projection coefficients of transformed Hamiltonian terms, we provide an explicit matrix-vector formulation of the search problem, which translates directly to linear equality and inequality constraint formulation.  Because minimizing both the sequence length and cardinality is desirable in practice, the optimization objective function includes both of these quantities.  We explicitly show how to formulate the optimization for linear or integer programming.  We apply this formalism to the problem of dipolar decoupling for broadband magnetometry in spin ensembles.  The goal is to decouple inhomogeneous interactions while retaining susceptibility to an external physical quantity of interest in order to prolong coherence in a way that directly translates to improved sensitivity.  
This motivates the definition of success criteria for achievability - clean Zeeman, and optimality - Zeeman strength.  
The HoRD-qubit-5 pulse sequence for spin-1/2 qubits provides an intuitive example and comparison with the well-known WHH-4 sequence shows its utility.  
We apply the constrained optimization formalism using linear programming in IBM CPLEX to discover the HoRD-qutrit-8 pulse sequence for qutrit ensembles, which achieves a clean Zeeman average Hamiltonian with Zeeman strength $\beta = 1/3$, the same Zeeman strength as HoRD-qubit-5.  
This sequence can be used in both the single quantum and double quantum bases, making it compatible with existing common-mode noise rejection techniques and spin-bath decoupling protocols.  
This work represents the first time, to our knowledge, decoupling techniques compatible with Ramsey magnetometry have been presented and analyzed, extending multiple pulse sequences from AC to DC sensing.  
Additionally, the NV$^{\text -}$-NV$^{\text -}$ decoupling approach proposed here overcomes a previously-viewed fundamental limit posed by sensing spin interactions.  

\ack
We appreciate discussions with Lev Bishop and John Smolin, as well as insightful comments from Murphy Yuezhen Niu after a thorough reading of this manuscript.  
This material is based upon work supported by the MIT-IBM Watson AI Lab and MIT under Air Force Contract No. FA8702-15-D-0001. Any opinions, findings, conclusions or recommendations expressed in this material are those of the author(s) and do not necessarily reflect the views of the MIT.

\appendix
\section*{Appendix A
\label{sec:appendix_A}}
\setcounter{section}{1}

This section contains explicit expressions for the Pauli matrices, spin-1 operators, and Gell-Mann matrices, as well as useful mathematical properties and relationships.  

The Pauli matrices,
\begin{eqnarray}
\sigma_1 = \left( \begin{array}{cc}
  {0} & {1} \\
  {1} & {0} \\
  \end{array}
  \right), \quad
\sigma_2 = \left( \begin{array}{cc}
  {0} & {-i} \\
  {i} & {0} \\
  \end{array}
  \right), \quad
\sigma_3 = \left( \begin{array}{cc}
  {1} & {0} \\
  {0} & {-1} \\
  \end{array}
  \right), \quad
\end{eqnarray}
together with the identity operator, $\sigma_0 = I_{2\times 2}$, form a linear basis (with real coefficients) for 2$\times$2 Hermitian matrices.  

The spin operators for $S = 1$ are
\begin{eqnarray}
\label{eq:paulis}
S_x = \frac{1}{\sqrt{2}} \left( \begin{array}{ccc}
  {0} & {1} & {0} \\
  {1} & {0} & {1} \\
  {0} & {1} & {0} \\
  \end{array}
  \right),\nonumber \\
S_y = \frac{1}{\sqrt{2}} \left( \begin{array}{ccc}
  {0} & {-i}& {0} \\
  {i} & {0} & {-i} \\
  {0} & {i} & {0} \\
  \end{array}
  \right), 
  \label{eq:spin_1_operators}\\
S_z = \left( \begin{array}{ccc}
  {1} & {0} & {0} \\
  {0} & {0} & {0} \\
  {0} & {0} & {-1} \nonumber \\
  \end{array}
  \right).  
\end{eqnarray}
The traceless and Hermitian Gell-Mann matrices,
\begin{eqnarray}
\lambda_1 &= \left( \begin{array}{ccc}
  {0} & {1} & {0} \\
  {1} & {0} & {0} \\
  {0} & {0} & {0} \\
  \end{array}
  \right),\
\lambda_2 = \left( \begin{array}{ccc}
  {0} & {0} & {0} \\
  {0} & {0} & {1} \\
  {0} & {1} & {0} \\
  \end{array}
  \right),\
\lambda_3 = \left( \begin{array}{ccc}
  {0} & {0} & {1} \\
  {0} & {0} & {0} \\
  {1} & {0} & {0} \\
  \end{array}
  \right), \nonumber
  \\
\lambda_4 &= \left( \begin{array}{ccc}
  {0} & {-i}& {0} \\
  {i} & {0} & {0} \\
  {0} & {0} & {0} \\
  \end{array}
  \right),\
\lambda_5 = \left( \begin{array}{ccc}
  {0} & {0} & {0} \\
  {0} & {0} & {-i}\\
  {0} & {i} & {0} \\
  \end{array}
  \right),
\\
\lambda_6 &= \left( \begin{array}{ccc}
  {0} & {0} & {-i}\\
  {0} & {0} & {0} \\
  {i} & {0} & {0} \\
  \end{array}
  \right),\
\lambda_7 = \left( \begin{array}{ccc}
  {1} & {0} & {0} \\
  {0} & {-1}& {0} \\
  {0} & {0} & {0} \\
  \end{array}
  \right),\
\lambda_8 = \frac{1}{\sqrt{3}} \left( \begin{array}{ccc}
  {1} & {0} & {0} \\
  {0} & {1} & {0} \\
  {0} & {0} & {-2}\\
  \end{array}
  \right)
  \nonumber
\end{eqnarray}
generalize the Pauli basis for $SU(2)$ to $SU(3)$.  These matrices, together with the identity, form a linear basis for 3$\times$3 Hermitian matrices.  Both the Pauli and Gell-Mann matrices have normalization
\begin{equation}
  \text{tr} (\lambda_\mu \lambda_\nu) = 2 \delta_{\mu\nu}
\end{equation}
and we choose $\lambda_0 = \sqrt{\frac23}I_{3\times 3}$ to satisfy this relation.  
$\lambda_7$ and $\lambda_8$ commute with each other.  

The spin-1 matrices are related to the Gell-Mann matrices by
\begin{eqnarray}
    S_x &= \frac{1}{\sqrt{2}}(\lambda_1+\lambda_2), \nonumber \\
    S_y &= \frac{1}{\sqrt{2}}(\lambda_4+\lambda_5), \label{eq:Sxyz_ito_GM}\\
    S_z &= \frac{1}{2}(\lambda_7+\sqrt{3}\lambda_8). \nonumber
\end{eqnarray}

There are three independent $SU(2)$ subalgebras, 
\begin{eqnarray}
  &\{ \lambda_1, \lambda_4, \lambda_7 \} \nonumber \\
  &\{ \lambda_2, \lambda_5, \zeta \}, \quad
      \zeta = \frac{1}{2}(\sqrt{3} \lambda_8 - \lambda_7) \label{eq:su2_subalgebra} \\
  &\{ \lambda_3, \lambda_6, \eta \}, \quad
        \eta = \frac{1}{2}(\sqrt{3} \lambda_8 + \lambda_7) \nonumber
\end{eqnarray}
corresponding to the $\{0, +1\}$, $\{0, -1\}$, and $\{-1, +1\}$ states, respectively.  

\section*{Appendix B}
\setcounter{section}{2}
\label{sec:appendix_B}
The unitary operators comprising the HoZD-qutrit sequence, discussed in~\sref{subsec:qutrit}, are 
\begin{equation}
\def\arraystretch{1.25}
\hspace{-5.5pt}
\begin{array}{l}
    U_0 = (V_4W_2)_1 (V_2W_2)_2 (V_3W_2)_3, \\
    U_1 = (V_0W_2)_1 (V_5W_3)_2 (V_3W_2)_3, \\
    U_2 = (V_4W_2)_1 (V_1W_1)_2 (V_3W_0)_3, \\
    U_3 = (V_0W_2)_1 (V_1W_3)_2 (V_3W_0)_3, \\
    U_4 = (V_1W_0)_1 (V_3W_0)_2 (V_2W_1)_3, \\
    U_5 = (V_0W_0)_1 (V_4W_0)_2 (V_5W_3)_3, \\
    U_6 = (V_0W_2)_1 (V_5W_0)_2 (V_0W_2)_3, \\
    U_7 = (V_0W_2)_1 (V_2W_2)_2 (V_0W_2)_3, \\
    U_8 = (V_1W_2)_1 (V_3W_1)_2 (V_0W_0)_3, \\
    U_9 = (V_3W_2)_1 (V_3W_3)_2 (V_0W_0)_3, \\
    U_{10} = (V_3W_0)_1 (V_4W_0)_2 (V_3W_1)_3, \\
    U_{11} = (V_4W_0)_1 (V_3W_0)_2 (V_4W_3)_3, 
\end{array}
\label{eq:qutrit-dd-12}
\end{equation}
with equal weights $w_k = 1$ for all $k$.      

\section*{References}
\bibliographystyle{iopart-num}
\bibliography{hameng}

\end{document}